\documentclass[pra,twocolumn,showpacs,superscriptaddress,10pt]{revtex4-1}
\usepackage[colorlinks,citecolor=blue,linkcolor=red,dvipdfm]{hyperref}
\usepackage{amsmath,amssymb,graphicx}
\usepackage{times}

\begin{document}

\title{One-dimensional solitons in fractional Schr\"{o}dinger equation with a spatially modulated nonlinearity: nonlinear lattice}

\author{Liangwei Zeng}
\affiliation{State Key Laboratory of Transient Optics and Photonics, Xi'an
Institute of Optics and Precision Mechanics of CAS, Xi'an 710119, China}
\affiliation{University of Chinese Academy of Sciences, Beijing 100084, China}

\author{Jianhua Zeng}
\email{\underline{zengjh@opt.ac.cn}}
\affiliation{State Key Laboratory of Transient Optics and Photonics, Xi'an
Institute of Optics and Precision Mechanics of CAS, Xi'an 710119, China}
\affiliation{University of Chinese Academy of Sciences, Beijing 100084, China}


\begin{abstract}

{The existence and stability of stable bright solitons in one-dimensional (1D) media with a spatially periodical modulated Kerr nonlinearity are demonstrated by means of the linear-stability analysis and in direct numerical simulations. The nonlinear potential landscape can balance the fractional-order diffraction and thus stabilizes the solitons, making the model unique and governed by the recently introduced fractional Schr\"{o}dinger equation with a self-focusing cubic nonlinear lattice. Both 1D fundamental and multipole solitons (in forms of dipole and tripole ones) are found, which occupy one or three cells of the nonlinear lattice respectively, depending on the soliton's power (intensity). We find that the profiles of the predicted soliton families are impacted intensely by the L\'{e}vy index $\alpha$ which denotes the level of fractional Laplacian, so does to their stability. The stabilization of soliton families is possible if $\alpha$  exceeds a threshold value, below which the balance between fractional-order diffraction and the spatially modulated focusing nonlinearity will be broken.}

\end{abstract}

\maketitle

\section{Introduction}

The exploration of the existence and possible stability mechanism of solitons (solitary waves) is a subject that has been received broad and renewed interest---particularly in nonlinear optical media and Bose-Einstein condensates (BECs) \cite{soliton-rev1,soliton-rev2}.  The interaction of the diffraction/dispersion and nonlinearity can reach a balance and results into the formation of localized states. In one-dimensional (1D) cubic (Kerr) nonlinear media, the focusing nonlinearity is responsible for creating bright localized states (bright solitons). In multidimensional spaces, the stabilization of multidimensional solitons is an on-going challenging issue \cite{soliton-rev1,soliton-rev2}.  Despite stable solitons can be existed in 1D purely uniform nonlinear media, their stability region is restricted to limited physical parameter domains. In this, nonuniform nonlinear media are usually being deliberately taken into consideration. A popular way is to add some certain linear potentials, which make it possible to stabilize various kinds of solitons in any dimension (from 1D to three-dimensional settings), relying on external periodic potentials such as optical lattices in atomic BECs \cite{OL-rev}, photonic crystals \cite{PC-rev} and lattices \cite{PL-rev1,PL-rev2} in nonlinear optical media; we call such way the linear mechanism. Another different stabilization scheme can be viewed as nonlinear mechanism since it is possible only when adding inhomogeneous potentials to the purely nonlinear media in two ways: with spatially inhomogeneous defocusing nonlinearity (which was introduced several years ago and are receiving increasing attention)  \cite{defo1,defo2,defo5,defo6,defo7} and with a periodic modulation of nonlinearity \cite{soliton-rev1,CQ1D,CQ2D,SQ1D} , this last case is also known as nonlinear lattices. The combined linear-nonlinear mechanism \cite{LNL1D,LNL2D}, with a couple of linear lattices and nonlinear ones, was elaborated and can support for various types of stable solitons too in past years.

Above schemes to the stabilization of solitons are all within the territory of standard quantum mechanics, which can be extended to non-Hermitian system provided that its Hamiltonian is parity-time ($\mathcal{PT}$) symmetric, such system belongs to non-Hermitian quantum mechanics. Because of having $\mathcal{PT}$ symmetry, the non-Hermitian Hamiltonian---counterintuitively---exhibits entirely real eigenvalue spectra. Various nonlinear waves and their propagation dynamics in diverse $\mathcal{PT}$-symmetric systems have been a hot topic in the past decade, with a particular interest in those systems with periodic potentials---$\mathcal{PT}$ lattices, see a review \cite{PT-rev} and references therein. In addition to this, a different extension of the standard quantum mechanics has also been introduced to the system whose diffraction is a fractional order derivative of space instead of second derivative in its standard counterpart, such system is called space-fractional quantum mechanics (SFQM)---the name was coined by Nick Laskin in 2000 \cite{Lask1, Lask2} and derived from replacing the Brownian trajectories in Feynman path integrals by the L\'{e}vy flights. Just as the standard quantum mechanics is described by the Schr\"{o}dinger equation,  the fractional Schr\"{o}dinger equation (named by Laskin too \cite{Lask2, Lask3}) is the fundamental equation of physics for describing the SFQM.

Fractional Schr\"{o}dinger equation is recently received warmly welcome in many fields of physics, such as quantum physics \cite{Lask1, Lask2,Frac1}, condensed-matter physics \cite{Frac-condensed} and optics \cite{Frac-OL} , etc. Remarkably, an optical implementation of the fractional Schr\"{o}dinger equation, which is based on a spherical optical resonator that fulfills the fractional quantum harmonic oscillator, was proposed by Longhi in 2015 \cite{Frac-OL} . Such an initiative work, which paved a way toward studying the propagation dynamics of spatial optical beams in the field of optics to emulate the fractional models (such as the SFQM mentioned above) stemmed from quantum physics, is being followed up by blooming developments in various settings \cite{Frac2, Frac3, Frac4, Frac5, Frac6, Frac7, Frac8, Frac9, Frac10, Frac11, Frac12, Frac13, Frac14}.

Though the studies of solitons in the nonlinear fractional Schr\"{o}dinger equation (NLFSE) have made broad and rapid progress, their study in purely nonlinear setting integrated with a nonlinear lattice is lacking. To do that, the objective of this Letter is to assess the existence and stability property of 1D solitons in the fractional model combined the NLFSE and a focusing nonlinear lattice, numerically and theoretically. Noteworthy results are obtained. We find that such model, under different initial conditions (solitons' power), gives rise to abundant stable soliton family solutions, including 1D fundamental solitons and multipole ones of forms of dipole and tripole solitons, verified by linear-stability analysis to compute complex eigenvalues associated with each family of the 1D solitons and by direct simulations of their dynamics under sufficiently small initial perturbations. Profiles of the soliton families, together with their stability, are found to be deeply affected by the fractional-order diffraction (of the physical model), which is characterized by L\'{e}vy index $\alpha$. Besides, a threshold value of $\alpha$, below which all the numerically found soliton families are unstable, is identified.

\begin{figure}[tbp]
\begin{center}
\includegraphics[width=1\columnwidth]{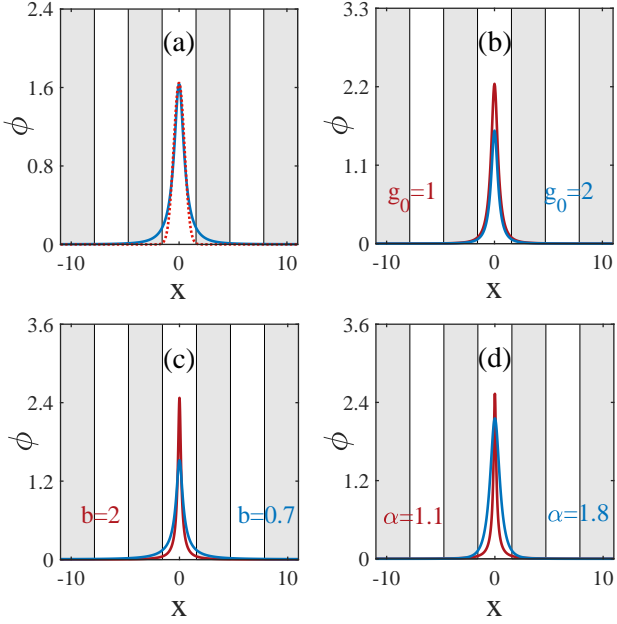}
\end{center}
\caption{(a) Profile (blue) of 1D fundamental soliton (number of nodes $k=0$), found in the fractional model  with $b=1$ and $\alpha=1.5$, and its comparison to the normal Gaussian solution (magenta dashed) given by Eq. (\ref{Ugaussian}) at $A_\mathrm{g}=1.65$ and $W=0.5$. Profiles of fundamental solitons for different values of the nonlinearity strength $\mathrm{g}_0$ (b), of the propagation constant $b$ (c), and of the L\'{e}vy index $\alpha$  (d). Other parameters are $b=2$ for panels (b) and (d), $\alpha=1.5$ for (b) and $\alpha=1.15$ for (c). We set $\mathrm{g}_0=1$ through out this letter except for that in Fig. \ref{fig1}(b). Alternating shaded and blank regions here and in the Fig. \ref{fig4} below are shown for lattice periods of the local nonlinearity in Eq. (\ref{gprofile}).}
\label{fig1}
\end{figure}

\begin{figure}[tbp]
\begin{center}
\includegraphics[width=1\columnwidth]{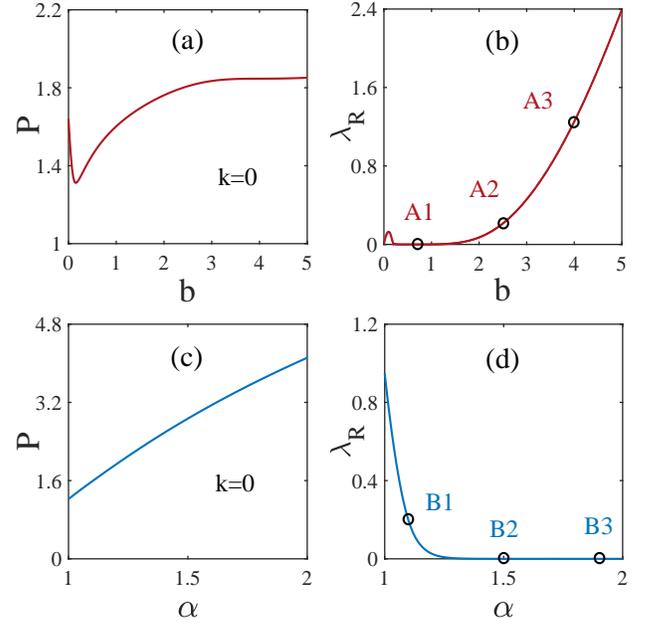}
\end{center}
\caption{(a) Fundamental soliton power $P$ and the relevant linear-stability analysis results (b) shown as maximal real part of eigenvalues $\lambda_{\rm R}$ [obtained numerically from Eq. (\ref{LAS})] versus propagation constant $b$ at $\alpha=1.15$. (c) $P$ and (d) $\lambda_{\rm R}$ versus $\alpha$ at $b=2$. The propagation simulations of the marked points (A1, A2, A3) and (B1, B2, B3) in panels (c) and (d), respectively, are displayed in Fig. \ref{fig3}.}
\label{fig2}
\end{figure}

\begin{figure}[tbp]
\begin{center}
\includegraphics[width=1\columnwidth]{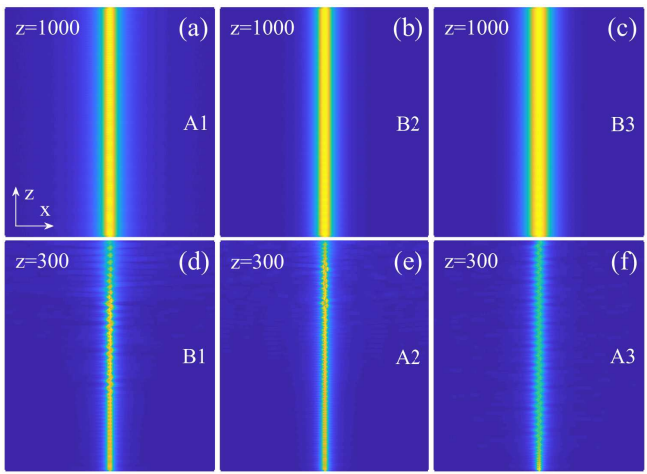}
\end{center}
\caption{Top row: stable propagations of the 1D fundamental solitons against small perturbations at $\alpha=1.15$, $b=0.7$ (a),  at $\alpha=1.5$, $b=2$ (b), and at $\alpha=1.9$, $b=2$ (c). Bottom row: unstable propagations of the perturbed 1D fundamental solitons at $\alpha=1.1$, $b=2$ (d), at $\alpha=1.15$, $b=2.5$ (e), and at $\alpha=1.15$, $b=4$ (f). $x\in[-6, 6]$ for all panels.}
\label{fig3}
\end{figure}

\section{Fractional model and numerical ways}
The propagation of paraxial laser beams under fractional-order diffraction is described by the NLFSE, which is written in the dimensionless form:
\begin{equation}
i\frac{\partial E}{\partial z}=\frac{1}{2}\left(-\frac{\partial^2}{\partial x^2}\right)^{\alpha/2}E-\mathrm{g}(x)\left|E\right|^2E.
\label{NLFSE}
\end{equation}
Here $E$ and $z$ denote the field amplitude and propagation distance respectively, while $\mathrm{g}(x)>0$ is the strength of self-focusing nonlinearity . Note that $(-\partial^2/\partial x^2)^{\alpha/2}$ represents the fractional Laplacian where $\alpha$ ($1<\alpha\leq2$) being the L\'{e}vy index. Particularly, Eq. (\ref{NLFSE})  degenerates to the well-know nonlinear Schr\"{o}dinger equation when $\alpha=2$. In this Letter, we define the form of periodic nonlinearity (nonlinear lattice) as
\begin{equation}
\mathrm{g}(x)=\mathrm{g_0}{\rm cos}^2(x).
\label{gprofile}
\end{equation}

We search the stationary field amplitude $U$ by $E=U~{\rm exp}(ibz)$ ($b$ is the propagation constant), which yields:
\begin{equation}
-bU=\frac{1}{2}\left(-\frac{\partial^2}{\partial x^2}\right)^{\alpha/2}U+\mathrm{g}(x)\left|U\right|^2U.
\label{NLFSES}
\end{equation}
The soliton power is given by $P=\int\left|U(x)\right|^2dx$. By solving Eq. (\ref{NLFSES}) with Gaussian beams as initial input , we can get the stationary solutions of solitons. We also compare the soliton solutions to the normal Gaussian function, which follows:
\begin{equation}
U_\mathrm{g}=A_\mathrm{g}{\rm exp}[-x^2/(2W^2)].
\label{Ugaussian}
\end{equation}
Here $A_\mathrm{g}$, $W$ (constant parameters) are the relevant amplitude and width, respectively.

The stability of the numerical found soliton solutions is a key issue. To this, we take the linear stability analysis method, by perturbing the field amplitude as $U=[U(x)+p(x){\rm exp}(\lambda z)+q^*(x){\rm exp}(\lambda ^*z)]{\rm exp}(ibz)$, where $U(x)$ is undisturbed field amplitude, $p(x)$ and $q^*(x)$ are small perturbations. This leads to the following eigenvalue equations from Eq. (\ref{NLFSE}):
\begin{equation}
\left\{
\begin{aligned}
i\lambda p=+\frac{1}{2}\left(-\frac{\partial^2}{\partial x^2}\right)^{\alpha/2}p+bp+\mathrm{g}U^2(2p+q),\\
i\lambda q=-\frac{1}{2}\left(-\frac{\partial^2}{\partial x^2}\right)^{\alpha/2}q-bq-\mathrm{g}U^2(2q+p).
\end{aligned}
\label{LAS}
\right.
\end{equation}
If the real parts of the eigenvalues, $\lambda_{\rm R}$, calculated by the eigenvalue problem (\ref{LAS}), are all zero (that is $\lambda_{\rm R}=0$), the perturbed solitons are stable. Otherwise, they are unstable solutions. The stationary states of Eq. (\ref{NLFSES}) are calculated by the modified squared-operator method \cite{MSOM}, and their propagation dynamics are tested by linear-stability analysis [Eq. (\ref{LAS})] and in direct numerical simulations of Eq. (\ref{NLFSE}) based on the well-known split-step Fourier method.

\section{Numerical results}
\subsection{Fundamental solitons}

The first soliton family we are interested in is the fundamental solitons [the ones have number of nodes (zeros) $k=0$] that have a single peak, since in a general nonlinear physical model ground states is an important role they usually play. One is legitimate to wonder about the issues like what are the conditions that such fundamental modes can exist, and how they can be made stable, and what are the differences between the ones generated in the considered fractional model and that from conventional model with second-order diffraction.

The above issues are partially addressed in Fig. \ref{fig1}, where shows the fundamental solitons under various constraint conditions with good convergence can be obtained numerically. Fig. \ref{fig1}(a) depicts the profiles of the numerical found solution (blue solid line) and the normal Gaussian solution (magenta dashed line) with the same soliton's power and amplitude. One can see from the panel that the fundamental soliton starts to expand at the waist (compared to its Gaussian counterpart), with decaying tails on both sides penetrating into nearest adjacent cells, keeping its main lobe within a single cell of the lattice, and resting on the lattice maximum. A comparison of their profiles with different nonlinear strength $\mathrm{g}_0$ reveals that both amplitude and width of the fundamental soliton decreases when increasing $\mathrm{g}_0$, according to Fig. \ref{fig1}(b). Under different propagation constants $b$, as displayed in Fig. \ref{fig1}(c), the soliton's amplitude increases while its waist (width) shrinks with an increase of $b$. Conversely, Fig. \ref{fig1}(d) demonstrates that the fundamental soliton expands and gets shorter with an increase of L\'{e}vy index $\alpha$ that denotes diffraction order.

\begin{figure}[tbp]
\begin{center}
\includegraphics[width=1\columnwidth]{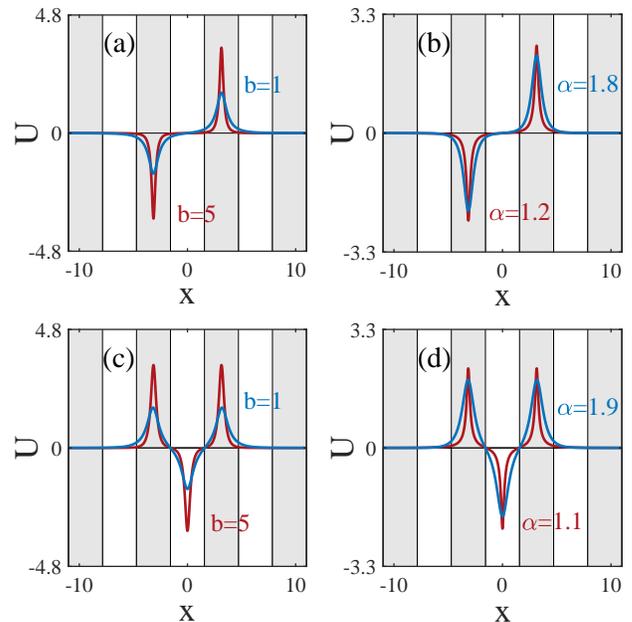}
\end{center}
\caption{Top row: profiles of dipole solitons (number of nodes $k=1$) for different values of the nonlinearity strength $\mathrm{g}_0$ at $\alpha=1.5$ (a) and of the L\'{e}vy index $\alpha$ at $b=2$ (b). Bottom row: the same but for tripole solitons with different values of $b$ at $\alpha=1.7$ (c) and of $\alpha$ at $b=1.5$ (d).}
\label{fig4}
\end{figure}

\begin{figure}[tbp]
\begin{center}
\includegraphics[width=1\columnwidth]{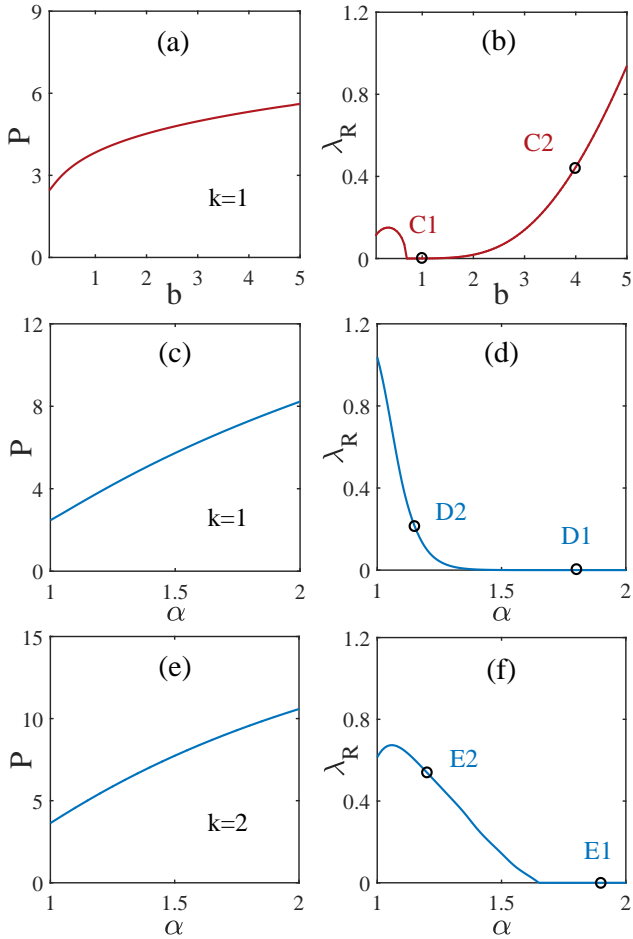}
\end{center}
\caption{(a) Dipole soliton power $P$ and (b) maximal real part of eigenvalues $\lambda_{\rm R}$ [given by Eq. (\ref{LAS})] versus propagation constant $b$ at L\'{e}vy index $\alpha=1.3$. (c) $P$ and (d) $\lambda_{\rm R}$ versus $\alpha$ for dipole solitons (number of nodes $k=1$) at $b=2$. (e) $P$ and (f) $\lambda_{\rm R}$ versus $\alpha$ for tripole solitons (number of nodes $k=2$) at $b=1.5$. The propagation simulations of the marked points (C1, D1, E1) and (C2, D2, E2) in panels (b), (d) and (f) are portrayed in the top and bottom rows of Fig. \ref{fig6}, respectively.}
\label{fig5}
\end{figure}

We then investigate two possible binding dependencies of the power $P$ of the fundamental soliton family at fixed L\'{e}vy index $\alpha$ (while changing $b$), and at fixed propagation constant $b$ (while changing $\alpha$), which are portrayed respectively as the curves  $P(b)$  and $P(\alpha)$ in Figs. \ref{fig2}(a) and \ref{fig2}(c). One can see from the former panel that $P$ initially decreases then increases with gradually increasing $b$, while the $P$  has linear relationship with the $\alpha$ for the latter panel---the power $P$  largens linearly when the L\'{e}vy index $\alpha$  increases. Linear stability analysis for both cases, in the forms of $\lambda_{\rm R}(b)$ and $\lambda_{\rm R}(\alpha)$ [$\lambda_{\rm R}$ is the maximal real part of the eigenvalues of the so found solutions], are presented in Figs. \ref{fig2}(b) and \ref{fig2}(d), which show that fundamental solitons can be completely stable modes provided that the relevant propagation constant $b$ is within certain domain (not too small and not too big), that the L\'{e}vy index $\alpha$ above a threshold value---its existence may be explained by the breaking balance between the nonlinearity and fractional-order diffraction (at small $\alpha$). These linear-stability results are confirmed with numerical simulations, as depicted in Fig. \ref{fig3}, from which we can see that the stable solitons keep their coherence over long propagation distance, while unstable solitons transform themselves into regular oscillating modes like breathers.

\subsection{Multipole solutions: dipole and tripole solitons}

Besides the fundamental solitonic modes, the considered fractional model---the Eq. (\ref{NLFSE}) with a nonlinear lattice given by  (\ref{gprofile})---also supports a vast variety of high-order localized modes, e.g., the solitons with different $k$ (number of nodes/zeros).
Examples of these modes appearing as dipole solitons and tripole modes are displayed in the top and bottom rows of Fig. \ref{fig4}, where clearly shows that the spacing $\Delta_s$ between adjacent solitons is $\Delta_s=2\pi$ (twice to that of the nonlinear lattice $\Delta_{NL}=\pi$) for the former modes, and $\Delta_s=\pi$ (equal to the nonlinear lattice period) for the latter. Systematic simulations of their propagations in Eq. (\ref{NLFSE}) confirm that the dipole solitons can be stable modes only if $\Delta_s\geq\pi$, below such value ($\pi$ ) repulsive interaction between each soliton destabilize their coherence; with regard to the tripole modes, identical solitons on both sides can balance the $\pi$ phase shift for the central soliton thus keep their coherence as a bound state, making them stable localized modes. The profiles of these modes under different scenarios, such as fixing L\'{e}vy index $\alpha$ (while varying $b$) and fixing propagation constant $b$ (while varying $\alpha$), exhibit typical features to that of their fundamental counterparts presented in Fig. \ref{fig1}.

\begin{figure}[tbp]
\begin{center}
\includegraphics[width=1\columnwidth]{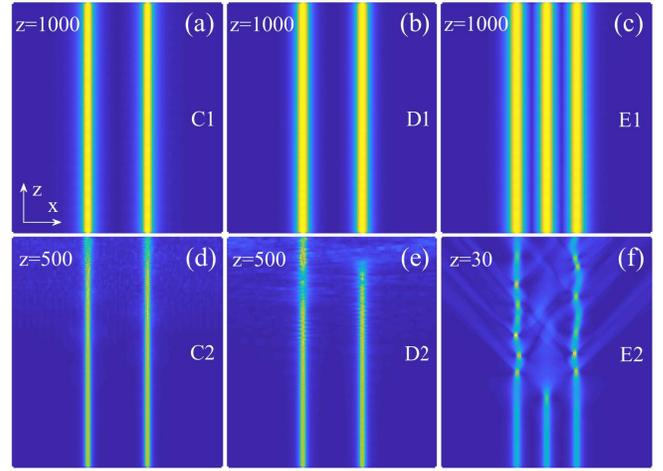}
\end{center}
\caption{Top row: stable propagations of 1D high-order (multipole) solitons against small perturbations for dipole solitons at $\alpha=1.3$, $b=1$ (a) and $\alpha=1.8$, $b=2$ (b), and for tripole soliton at $\alpha=1.9$, $b=1.5$ (c). Bottom row: unstable propagations of the perturbed 1D high-order solitons for dipole solitons at $\alpha=1.3$, $b=4$ (d) and $\alpha=1.15$, $b=2$ (e), and for tripole soliton at $\alpha=1.15$, $b=4$ (f).}
\label{fig6}
\end{figure}

The numerical found relation of $P$ versus $b$ for the dipole solitons (with number of zeros $k=1$) is displayed in Fig. \ref{fig5}(a),  it is seen that $P$ increases linearly with the increase of $b$ [we remark that the curve $P(b)$ for tripole solitons also obeys this trend, which is not shown, space limited]. It is naturally to know, from the Figs. \ref{fig5}(c) and \ref{fig5}(e), that the dependencies $P(\alpha)$ for dipole and tripole solitons follows this linear growth relation, resembling their fundamental counterpart in Fig. \ref{fig2}(c). Linear-stability analysis demonstrates that the stable multipole solitons at given $\alpha$ are restricted to limited region of $b$, just like that for fundamental ones, as compared the Fig. \ref{fig5}(b) with Fig. \ref{fig2}(b). With an increase of the number of solitons of which the soliton composites (multipole solitons) are composed, the stability regions for stable multipole modes at fixed $b$ reduce drastically, as prominently evidenced by their curves $\lambda_{\rm R}(\alpha)$ in Figs. \ref{fig5}(d) and \ref{fig5}(f). Systematic simulations performed
in stability and instability regions verified that the stable multipole solitons are robustly modes, while unstable ones either hold regular oscillations or lose their coherence completely and disappear as radiating waves during propagation, examples of both cases are displayed in the top and bottom lines of Fig. \ref{fig6}, respectively.

\section{Conclusion}
To summarize, we have studied the existence and stability of various 1D soliton families---including fundamental solitons and multipole ones (dipole and tripole solitons)---in the framework of nonlinear fractional Schr\"{o}dinger equation by combing the recently introduced  fractional Schr\"{o}dinger equation with a spatially periodic nonlinearity (nonlinear lattice) in the Kerr focusing optical or atomic media. Notably, interaction between the nonlinear potential landscape and the fractional-order diffraction can reach a balance and make the model is unique and suitable for stabilizing solitons, confirmed by linear stability analysis and direct simulations of the perturbed propagation. We found that the L\'{e}vy index $\alpha$ greatly affects the profiles of the thus found soliton families and their stability.  Stability and instability regions of all the soliton families under different parameter spaces were obtained. The results presented here are extended naturally to two-dimensional geometry to construct stable localized modes---solitons and vortices, which is an open issue by solely using the smoothly modulated nonlinear lattices.

\section{Acknowledgment}

\textbf{This} work was supported, in part, by the Natural Science Foundation of China (project Nos. 61690224, 61690222), by the Youth Innovation Promotion Association of the Chinese Academy of Sciences (project No. 2016357).

\end{document}